\documentclass[11pt,a4paper]{article}

\usepackage[margin=1in]{geometry}
\usepackage[T1]{fontenc}
\usepackage[utf8]{inputenc}
\usepackage{lmodern}
\usepackage{microtype}
\usepackage{amsmath,amssymb}
\usepackage{booktabs}
\usepackage{array}
\usepackage{tabularx}
\usepackage{threeparttable}
\usepackage{graphicx}
\usepackage{tikz}
\usetikzlibrary{arrows.meta,positioning,fit,backgrounds}
\usepackage{enumitem}
\usepackage{cite}
\usepackage{url}
\usepackage[hidelinks]{hyperref}
\usepackage{orcidlink}
\usepackage{float}

\renewcommand{\arraystretch}{1.08}

\usepackage{orcidlink}

\title{\textbf{Toward Decentralized Carbon Trading in Indonesia:\\
A Public-Blockchain Architecture for Tokenized Real-World Assets}}

\author{
\begin{minipage}[t]{0.31\textwidth}
\centering
\textbf{Rischan Mafrur\textsuperscript{*}}\,\orcidlink{YOUR-ORCID-ID}\\[0.3em]
\small Western Sydney University\\
\small Australia\\[0.3em]
\small \texttt{R.Mafrur@westernsydney.edu.au}
\end{minipage}
\hfill
\begin{minipage}[t]{0.31\textwidth}
\centering
\textbf{Fadli Ikhsan Pratama}\\[0.3em]
\small Indonesian Financial Services Authority (OJK)\\
\small Indonesia\\[0.3em]
\small \texttt{fadli.ikhsan@ojk.go.id}
\end{minipage}
\hfill
\begin{minipage}[t]{0.31\textwidth}
\centering
\textbf{Khadijah}\\[0.3em]
\small CryptocoinHalal.com\\
\small Indonesia\\[0.3em]
\small \texttt{khadijahasim1@gmail.com}
\end{minipage}
}

\begin{document}

\maketitle

\begin{abstract}
Indonesia has established a regulated carbon market supported by national registry infrastructure and the IDXCarbon exchange. Carbon units can be issued, recorded, traded, and retired within this framework. IDXCarbon currently uses a private blockchain for its trading infrastructure. This creates an opportunity to examine how Indonesian carbon credits could also be represented and traded through public blockchain infrastructure.

This study proposes an architecture for tokenizing Indonesian carbon credits as real-world assets (RWAs), with particular focus on \textit{Sertifikat Pengurangan Emisi Gas Rumah Kaca} (SPE-GRK). The proposed architecture retains the Sistem Registri Unit Karbon (SRUK) as the
authoritative source of carbon-unit status. It introduces a public-blockchain layer for token
representation and programmable transactions. The architecture is designed to support
lifecycle management, token-based asset representation, public observability of token activity,
interoperability, wallet-based transactions, and programmable settlement.

The architecture consists of four layers: the authoritative carbon layer, the registry interoperability and tokenization layer, the public-blockchain RWA layer, and the market and application layer. Access to the tokenized carbon assets remains regulated. Token issuance and transfers are linked to participant eligibility and registry status. Retirement also remains dependent on the authoritative carbon registry. The proposed architecture provides a framework for introducing public-blockchain RWA infrastructure into Indonesia's existing carbon market while maintaining SRUK authority and existing market-integrity controls.
\end{abstract}


\section{Introduction}
\label{sec:introduction}

Carbon markets are increasingly intersecting with developments in programmable financial infrastructure. Distributed ledgers, smart contracts, tokenized assets, and programmable settlement are no longer confined to cryptocurrency markets; they are increasingly being examined within regulated financial systems, including securities, banking, payments, and real-world asset (RWA) markets \cite{BIS2024,Agur2025,Adrian2026,BISAgora2026}. Carbon markets have followed a related trajectory. Prior research has examined blockchain-based registries, peer-to-peer carbon trading, smart-contract settlement, tokenized carbon credits, and public-chain carbon markets, with several models progressing beyond conceptual proposals to operational implementations \cite{VilkovTian2023,Sorensen2023,Swinkels2024,Ballesteros2024,Parhamfar2024,Abiodun2025}.

Carbon credits have already been tokenized and traded on blockchain infrastructure \cite{Sorensen2023,Swinkels2024}, public-chain carbon markets have already been observed empirically \cite{Ballesteros2024}, and blockchain architectures have been proposed for regulated emissions-trading systems \cite{Ajakwe2025}. At the same time, the emerging evidence cautions against treating tokenization as an automatic solution to carbon-market frictions. Tokenized markets can remain illiquid or concentrated \cite{Swinkels2024,Tlili2025,Mafrur2026}; public-chain infrastructures can reproduce new forms of intermediation and governance concentration \cite{Bassi2025}; and increasing financial composability can weaken the visibility of the relationship between a traded token and the environmental characteristics of the mitigation project that underpins it \cite{Tan2024}. The research problem is therefore not whether carbon credits \emph{can} be placed on a blockchain, but how they can be represented and transacted on public programmable infrastructure without weakening the institutional and environmental controls that establish their validity.

Indonesia provides a particularly relevant setting for examining this problem. Presidential Regulation No.\ 110 of 2025 (\textit{Peraturan Presiden Nomor 110 Tahun 2025}, hereafter Perpres 110/2025) establishes the current national framework for Carbon Economic Value (\textit{Nilai Ekonomi Karbon}, NEK) instruments and greenhouse-gas emission control \cite{Perpres110}. The framework recognizes carbon trading through a Carbon Exchange and direct trading and places national carbon-unit recording within the \textit{Sistem Registri Unit Karbon} (SRUK) framework \cite{Perpres110}. Carbon Exchange activity is regulated by the Financial Services Authority (\textit{Otoritas Jasa Keuangan}, OJK), while the Indonesia Stock Exchange operates IDXCarbon as the regulated carbon-exchange platform \cite{OJK14,OJK10}.

Indonesia already operates a digital carbon market through IDXCarbon. The platform uses blockchain technology and provides auction, regular continuous-auction, negotiated, and marketplace trading mechanisms \cite{IDXHome,IDXFAQ,IDXSPE}. Secondary technical and media sources characterize its blockchain implementation as private \cite{TruCarbon2026,BloombergTechnoz2024}. Accordingly, the contribution of this study does not lie in introducing blockchain to Indonesian carbon trading. The focus is on the additional capabilities that may arise when Indonesian carbon credits are represented as RWAs on public blockchain infrastructure.

Tokenization provides one potential source of such incremental functionality. An Indonesian
carbon unit such as the Sertifikat Pengurangan Emisi Gas Rumah Kaca (SPE-GRK) is already
digitally recorded and can already be traded. Tokenization introduces an additional standardized
representation on programmable infrastructure. Token holding, transfer restrictions, lifecycle
events, settlement conditions, and application interfaces can then be incorporated into
smart-contract logic \cite{BIS2024,Agur2025}. In RWA terms, the problem is one of representing
an already-recognized off-chain asset on a programmable ledger while preserving a reliable
relationship with the authoritative system that determines the asset's existence and status.

Public blockchain introduces a second and narrower design question. Private and consortium ledgers can also provide smart contracts, rapid settlement, and controlled participation. The potential incremental value of public infrastructure instead lies in providing a shared execution environment in which token-side state and smart-contract operations can be independently observed and verified beyond the infrastructure of a single market operator \cite{NIST8202,NIST8301}. Public deployment can also provide standardized interfaces to wallets, token infrastructure, and compatible applications. These properties may increase asset portability and programmability across authorized financial, corporate, accounting, and supervisory applications.

However, public infrastructure should not be conflated with permissionless carbon-market participation. A public blockchain may provide open transaction validation and observable contract state while the carbon asset itself remains subject to controlled issuance, verified identities, transfer restrictions, regulatory supervision, and authoritative registry status. More fundamentally, blockchain consensus cannot establish whether a mitigation activity produced a valid carbon unit. Environmental validity remains dependent on methodology, measurement, reporting and verification (MRV), institutional authorization, and sovereign registry governance \cite{VilkovTian2023,Tan2024}. The relevant design challenge is therefore to decentralize selected aspects of asset representation and transaction execution without decentralizing the legal and institutional authority over carbon validity. In this study, decentralization refers to public-blockchain
validation and programmable execution of the tokenized representation. It does not extend
to carbon-unit issuance, environmental validity, participant eligibility, or regulatory authority.

This challenge reveals a gap between two streams of existing research. Studies of blockchain within regulated emissions-trading systems have predominantly emphasized permissioned or hybrid architectures that preserve institutional control \cite{Ajakwe2025,Abiodun2025}. By contrast, empirical studies of public-chain carbon tokenization have largely examined voluntary carbon markets and crypto-financial ecosystems such as AirCarbon Exchange and KlimaDAO \cite{Swinkels2024,Ballesteros2024,Bassi2025}. Less attention has been given to the architectural problem that arises when a carbon unit governed by a sovereign national registry is represented as an RWA on public blockchain infrastructure while registry primacy, provenance, participant eligibility, lifecycle synchronization, and authoritative retirement must all be preserved.

This study addresses this gap in the Indonesian carbon-market context. It develops a public-blockchain RWA architecture in which SRUK remains the authoritative source of carbon-unit status. Public blockchain is introduced as an additional layer for token representation and programmable transactions. The architecture is designed to integrate with existing carbon-market institutions, including SRUK and IDXCarbon.

Accordingly, the study addresses the following research questions:

\begin{description}[leftmargin=1.2cm,labelwidth=0.9cm,style=nextline]

\item[\textbf{RQ1}]
What incremental benefits and challenges can public-blockchain tokenization provide relative to Indonesia's existing digital and blockchain-enabled carbon-market infrastructure?

\item[\textbf{RQ2}]
What architecture is required to represent Indonesian carbon credits as RWAs on public blockchain infrastructure while preserving authoritative registry control and enabling regulated programmable trading?

\item[\textbf{RQ3}]
What technical, legal and regulatory, governance, institutional, and market conditions must be addressed for such an architecture to be implemented in Indonesia?

\end{description}

The study makes three contributions. First, it evaluates the incremental role of public-chain
tokenization within Indonesia's existing digital and regulated carbon market. This focuses the
analysis on functions that are not already provided by the current infrastructure. Second, it
develops a registry-anchored architecture for representing Indonesian carbon credits as
public-chain RWAs. The architecture maintains the link to SRUK and incorporates controls for
issuance, participant eligibility, provenance, lifecycle synchronization, and retirement. Third, the study identifies the technical, legal and regulatory, governance, institutional, and market conditions that would need to be addressed before such an architecture could be implemented in Indonesia.

\section{Indonesia's Carbon-Market Architecture and Current Market State}
\label{sec:indonesia_market}

This section establishes the institutional and market baseline against which the proposed public-blockchain RWA architecture is evaluated. This baseline is important because Indonesia does not represent a greenfield case in which blockchain technology is being introduced into an otherwise non-digital carbon market. Carbon units are already issued within a sovereign regulatory framework, recorded through national registry infrastructure, and transacted through regulated market channels. The relevant question is therefore whether a public-chain RWA layer can provide incremental functionality without displacing the institutions that establish carbon-unit validity.

\subsection{Institutional and Regulatory Architecture}
\label{subsec:institutional_architecture}

Indonesia's carbon market operates within the broader \textit{Nilai Ekonomi Karbon} (NEK), or Carbon Economic Value, framework established under Presidential Regulation No.\ 110 of 2025 (Perpres 110/2025) \cite{Perpres110}. The framework distinguishes the processes through which carbon units are generated and recognized from the mechanisms through which those units may subsequently be transacted. This distinction is fundamental to the present study because the existence of a trading infrastructure does not itself determine whether an underlying mitigation outcome constitutes a valid carbon unit.

For project-based emission reductions, the mitigation activity precedes the creation of a tradable carbon unit. Measurement, reporting, validation, and verification provide the evidentiary basis for recognizing the claimed mitigation outcome. In the case of the \textit{Sertifikat Pengurangan Emisi Gas Rumah Kaca} (SPE-GRK), Perpres 110/2025 provides for registration within the applicable national system, independent verification, and the use of verified mitigation results as a basis for issuance \cite{Perpres110}. IDXCarbon similarly describes SPE-GRK as an emission-reduction certificate associated with a mitigation activity that has completed the relevant measurement, reporting, and verification process and has been assigned a registry identifier \cite{IDXSPE}.

For analytical purposes, the institutional lifecycle can be represented as

\[
\text{Mitigation Activity}
\rightarrow
\text{MRV and Verification}
\rightarrow
\text{SPE-GRK}
\rightarrow
\text{SRUK}
\rightarrow
\text{Trading}
\rightarrow
\text{Retirement}.
\]

The sequence is deliberately simplified. Its purpose is to distinguish the processes that establish and maintain the authoritative carbon unit from the market mechanisms through which that unit may subsequently change ownership or be retired.

Indonesian carbon trading is not confined to a single trading venue. Perpres 110/2025 recognizes transactions conducted through a Carbon Exchange as well as direct trading, while requiring carbon-market activity to remain connected to the applicable national carbon-unit registry framework \cite{Perpres110}. This institutional separation between the registry and the trading channel is particularly important for the architecture developed later in this paper: introducing an alternative technical execution layer does not, by itself, create a new category of carbon unit or remove transactions from the applicable regulatory perimeter.

Carbon Exchange activity is supervised by the Financial Services Authority (Otoritas Jasa
Keuangan, OJK). POJK No.\ 14 of 2023 established the regulatory framework for carbon
trading through a Carbon Exchange \cite{OJK14}. POJK No.\ 10 of 2026 subsequently
amended this framework to align it with Perpres 110/2025. Among the relevant changes,
carbon units traded through a Carbon Exchange are required to be recorded in the Sistem
Registri Unit Karbon (SRUK), replacing the earlier regulatory reference to SRN-PPI
\cite{OJK10}. OJK subsequently issued PADK No.\ 6 of 2026 concerning the procedures
for conducting carbon trading through a Carbon Exchange \cite{OJKPADK6}. These instruments
form the current regulatory basis for Carbon Exchange activity considered in this study.

Commercial platforms may also provide user-facing access to regulated carbon-market
infrastructure. Jejakin's terms of service state that SPE-GRK made available through its
platform are sourced and transacted through IDXCarbon \cite{JejakinTOS2026}. For these
SPE-GRK transactions, Jejakin functions as a facilitator for access, transaction processing,
and retirement coordination. This role should be distinguished from that of IDXCarbon as
the underlying Carbon Exchange.

\begin{figure}[H]
\centering

\resizebox{0.88\textwidth}{!}{
\begin{tikzpicture}[
    node distance=1.1cm and 1.4cm,
    box/.style={
        draw,
        rounded corners,
        align=center,
        minimum height=0.95cm,
        text width=2.9cm,
        inner sep=5pt
    },
    smallbox/.style={
        draw,
        rounded corners,
        align=center,
        minimum height=0.9cm,
        text width=2.5cm,
        inner sep=5pt
    },
    interfacebox/.style={
        draw,
        rounded corners,
        align=center,
        minimum height=0.9cm,
        text width=3.0cm,
        inner sep=5pt
    },
    authoritybox/.style={
        draw,
        rounded corners,
        align=center,
        minimum height=0.9cm,
        text width=3.0cm,
        inner sep=5pt
    },
    arrow/.style={
        -{Latex[length=2.2mm]},
        thick
    },
    biarrow/.style={
        {Latex[length=2.2mm]}-{Latex[length=2.2mm]},
        thick
    },
    supervision/.style={
        -{Latex[length=2.2mm]},
        thick,
        dashed
    }
]


\node[box] (project)
{Mitigation\\Project};

\node[box, right=of project] (mrv)
{Measurement, Reporting\\and Verification};

\node[box, right=of mrv] (spe)
{Carbon Unit\\e.g., SPE-GRK};

\node[box, right=of spe] (sruk)
{SRUK\\Authoritative Carbon-Unit Registry};


\node[authoritybox, above=1.2cm of sruk] (authority)
{KLH/BPLH /\\Carbon-Market Authority};


\node[smallbox, below left=1.8cm and -0.1cm of sruk] (direct)
{Direct\\Trading};

\node[smallbox, below right=1.8cm and -0.1cm of sruk] (exchange)
{Carbon Exchange\\e.g., IDXCarbon};


\node[smallbox, below=1.4cm of direct] (parties)
{Buyer / Seller};


\node[smallbox, below left=1.4cm and -0.1cm of exchange]
(exchangeparticipants)
{Registered Exchange\\Participants};

\node[interfacebox, below right=1.4cm and -0.1cm of exchange]
(jejakin)
{Commercial Access Platform\\
e.g., Jejakin};


\node[smallbox, below=1.25cm of jejakin] (endusers)
{Corporate /\\Individual Buyers};


\node[authoritybox, right=1.8cm of exchange] (ojk)
{OJK\\Regulation and\\Supervision};


\draw[arrow] (project) -- (mrv);
\draw[arrow] (mrv) -- (spe);
\draw[arrow] (spe) -- (sruk);


\draw[supervision]
(authority.south) -- (sruk.north);


\draw[biarrow]
(sruk.south west)
-- node[
    left,
    midway,
    font=\footnotesize,
    align=center
]
{transaction\\recording}
(direct.north);

\draw[biarrow]
(sruk.south east)
-- node[
    right,
    midway,
    font=\footnotesize,
    align=center
]
{transaction\\recording}
(exchange.north);


\draw[biarrow]
(direct) -- (parties);


\draw[biarrow]
(exchange) -- (exchangeparticipants);

%

\draw[biarrow]
(exchange) -- (jejakin);

\draw[biarrow]
(jejakin) -- (endusers);


\draw[supervision]
(ojk.west) -- (exchange.east);

\end{tikzpicture}
}

\caption{Simplified institutional structure of Indonesian carbon trading. Carbon transactions
may occur through direct trading or a regulated Carbon Exchange, with transaction information
recorded through SRUK. Jejakin illustrates a commercial facilitation interface for SPE-GRK
transactions sourced and transacted through IDXCarbon.}
\label{fig:currentarchitecture}

\end{figure}
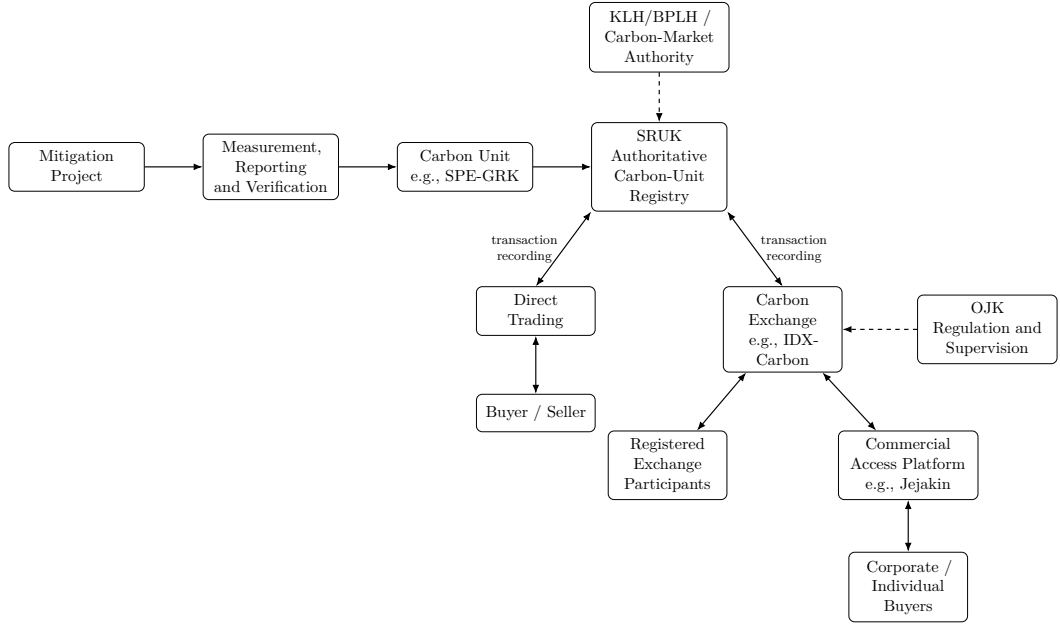

Figure~\ref{fig:currentarchitecture} summarizes this institutional structure. The figure
distinguishes the processes that establish the carbon unit and its authoritative registry state
from the channels through which the unit may be traded.


The distinction illustrated in Figure~\ref{fig:currentarchitecture} establishes an important boundary for the present study. MRV, verification, recognition of eligible carbon units, registry status, and regulatory supervision remain institutional functions. Accordingly, decentralization at the transaction or execution layer should not be interpreted as decentralization of the authority to determine whether a mitigation outcome constitutes a valid Indonesian carbon unit.

\subsection{SRUK and Authoritative Carbon-Unit State}
\label{subsec:sruk}

The transition from SRN-PPI to SRUK should be interpreted in the context of recent regulatory
changes. SRUK was officially launched by KLH/BPLH on 9 July 2026 as part of Indonesia's
updated national carbon-market infrastructure \cite{SRUKLaunch2026}. Its operational
implementation was therefore still at an early stage at the regulatory cut-off used in this study. Some IDXCarbon materials continue to refer to SRN-PPI when describing registry integration
\cite{IDXFAQ,IDXSPE}. POJK No.\ 10 of 2026 adopts SRUK within the updated Carbon
Exchange regulatory framework \cite{OJK10}. The two terms therefore reflect different stages
of Indonesia's registry transition and should not be used interchangeably.

For the purposes of RWA tokenization, the relevant institutional principle is independent of
the technology used by the registry. The sovereign registry remains the authoritative source
of carbon-unit state. At minimum, a tokenization arrangement would need to resolve those attributes necessary to establish the identity and status of the represented carbon unit, including its registry or batch identifier, underlying project, quantity, vintage, applicable methodology or standard, current status, and retirement state. This does not imply that all such information should be stored publicly. Commercially sensitive, personal, or legally restricted information can remain within authoritative or permissioned systems, while the public representation exposes only the identifiers, attestations, or cryptographic commitments required to establish provenance and reconcile asset state.

SRUK and the public blockchain therefore perform different functions. SRUK establishes
the authoritative status of the carbon unit. A blockchain can execute predefined rules using
authenticated registry information, but it cannot independently determine whether the underlying
emission reduction occurred, whether the applicable methodology was correctly applied, or
whether Indonesian authorities continue to recognize the carbon unit.

\subsection{IDXCarbon as the Existing Digital Trading Layer}
\label{subsec:idxcarbon}

IDXCarbon establishes that Indonesia already possesses an operational blockchain-enabled
carbon-trading venue. The platform provides auction, regular continuous-auction, negotiated,
and marketplace trading mechanisms \cite{IDXHome,IDXFAQ}. SPE-GRK may be traded at the
project level through auction, marketplace, and negotiated mechanisms, with eligible standardized
groupings also available through the regular market \cite{IDXSPE}. IDXCarbon's public documentation confirms the use of blockchain technology but does not
specify the underlying network, validator structure, or consensus arrangement. Secondary
reporting identifies the trading system as being built on a private blockchain with instantaneous
settlement \cite{BloombergTechnoz2024}.


Under the existing structure, the asset remains within institution-specific registry
and trading infrastructure. Public-chain tokenization would introduce an additional standardized
on-chain representation that can be independently inspected and programmatically accessed
across authorized applications. This provides the baseline for evaluating whether public-chain
tokenization offers functionality that is incremental to Indonesia's existing infrastructure.

\subsection{Current IDXCarbon Market Evidence}
\label{subsec:idxcarbon_market}

Table~\ref{tab:idxcarbon_monthly} reports monthly IDXCarbon market indicators for September 2025
through July 2026. These observations describe activity reported through IDXCarbon and do not
represent all carbon transactions conducted through other permitted channels, including direct
trading. Because the eleven-month observation window spans two calendar years, its aggregate
values should not be interpreted as either an official calendar-year total or an official
year-to-date statistic.

\begin{table}[htbp]
\centering
\begin{threeparttable}

\caption{Monthly IDXCarbon market indicators, September 2025--July 2026}
\label{tab:idxcarbon_monthly}

\small
\setlength{\tabcolsep}{4pt}
\renewcommand{\arraystretch}{1.05}

\begin{tabular}{@{}l r r c c r@{}}
\toprule
\textbf{Period} &
\textbf{Trading Volume} &
\textbf{Trading Value} &
\textbf{Listed} &
\textbf{Market} &
\textbf{Retired Units}\\

&
\textbf{(tCO$_2$e)} &
\textbf{(IDR)} &
\textbf{Projects} &
\textbf{Participants} &
\textbf{(tCO$_2$e)}\\
\midrule

Sep 2025 & 1,234   & 82,736,000     & 8  & 132 & 2,983   \\
Oct 2025 & 601     & 37,753,600     & 8  & 137 & 1,330   \\
Nov 2025 & 15,012  & 1,015,714,200  & 9  & 145 & 1,261   \\
Dec 2025 & 190,264 & 7,486,216,776  & 9  & 150 & 13,854  \\
Jan 2026 & 117,455 & 4,701,187,600  & 9  & 152 & 28,595  \\
Feb 2026 & 2,218   & 164,176,400    & 10 & 153 & 11      \\
Mar 2026 & 43,117  & 1,839,392,200  & 10 & 153 & 302,982 \\
Apr 2026 & 554     & 42,581,600     & 10 & 155 & 181     \\
May 2026 & 219     & 13,338,600     & 10 & 155 & 255     \\
Jun 2026 & 656     & 47,722,800     & 10 & 155 & 1,965   \\
Jul 2026 & 1,021   & 74,366,400     & 10 & 155 & 0       \\

\midrule

\textbf{Sep 2025--Jul 2026 sum} &
\textbf{372,351} &
\textbf{15,505,186,176} &
-- &
-- &
\textbf{353,417}\\

\midrule

\textbf{Official 2025 year-end} &
\textbf{903,915} &
\textbf{36,365,999,176} &
\textbf{9} &
\textbf{150} &
--\\

\textbf{Official 2026 YTD, Jul} &
\textbf{165,240} &
\textbf{6,882,765,600} &
\textbf{10} &
\textbf{155} &
--\\

\bottomrule
\end{tabular}

\begin{tablenotes}[flushleft]
\footnotesize
\item \textit{Notes:} Monthly figures are compiled from IDXCarbon monthly reports
\cite{IDXMonthly}. The ``Sep 2025--Jul 2026 sum'' is calculated from the
eleven monthly observations shown in the table. It should not be interpreted
as an official calendar-year or year-to-date figure. The official 2025
year-end and July 2026 YTD figures are included separately for reference.
Trading volume and retired carbon units are reported as separate indicators
by IDXCarbon.
\end{tablenotes}

\end{threeparttable}
\end{table}

The market data provide several observations relevant to the subsequent analysis. First, Indonesia already has an operational carbon market. Second, the number of listed SPE-GRK projects increased from eight in September 2025 to ten by February 2026, and the number of registered participants increased from 132 to 155 over the observation period \cite{IDXMonthly}. These developments indicate gradual market expansion, although the number of listed projects remains limited. Third, trading activity is highly uneven across months. Of the 165,240 tCO$_2$e reported for January--July 2026, January and March together account for the overwhelming majority of the observed volume. This concentration indicates that aggregate transaction volume alone does not necessarily imply continuous secondary-market activity.

The composition of trading activity reinforces this observation. Of the 165,240 tCO$_2$e reported during January--July 2026, approximately 160,882 tCO$_2$e was executed through negotiated transactions, compared with 4,099 tCO$_2$e through the marketplace and 259 tCO$_2$e through the regular market; no auction volume was reported over the same period \cite{IDXMonthly}. Negotiated transactions therefore accounted for approximately 97.4\% of reported 2026 year-to-date volume through July. This concentration should not be interpreted as evidence that a different technological architecture would necessarily produce greater liquidity. Empirical studies of tokenized carbon and other RWA markets show that the creation of a transferable on-chain representation does not by itself generate active secondary trading \cite{Swinkels2024,Tlili2025,Mafrur2026}. Liquidity remains dependent on the availability and quality of the underlying assets, buyer and seller participation, information, pricing, market-making incentives, transaction costs, and regulatory structure.

The IDXCarbon data provide a baseline for evaluating the proposed architecture. The analysis focuses on whether a public-chain RWA model can add capabilities that are not available in the current market structure, including standardized asset interfaces, programmable lifecycle functions, authorized wallet-based transfers, and independently observable token activity. Any effect on market participation or liquidity would require separate empirical evaluation.

\section{Literature Review}
\label{sec:literature}

\subsection{Blockchain as Carbon-Market Infrastructure}

The application of blockchain to carbon markets is no longer a predominantly conceptual research topic \cite{VilkovTian2023,Abiodun2025}. Within the literature, blockchain is principally treated as an infrastructure for maintaining auditable transaction records, coordinating distributed participants, and executing predefined market rules through smart contracts. Parhamfar et al.\ further identify peer-to-peer trading and automated settlement as recurring features of blockchain-enabled carbon-market designs, while also emphasizing continuing challenges related to scalability, interoperability, governance, and regulation \cite{Parhamfar2024}.

More recent work demonstrates that these concepts can also be embedded within regulated carbon-market structures. Ajakwe et al.\ develop and experimentally evaluate a hybrid blockchain architecture for the Korean Emissions Trading Scheme (K-ETS), integrating emissions data, regulator-controlled validation, allowance administration, trading, and compliance processes \cite{Ajakwe2025}. Golshani and Mankar similarly propose an architecture in which carbon-credit issuance and lifecycle events are recorded on blockchain while certification and selected market functions remain external to the ledger \cite{GolshaniMankar2026}. These studies are important because they demonstrate that blockchain-based transaction processing does not require the displacement of regulatory institutions.

The literature therefore provides substantial evidence that distributed ledgers and smart contracts can support carbon-market operations. Consequently, the relevant research question has shifted from whether blockchain can be used for carbon trading to how particular blockchain architectures should be integrated with the institutional, environmental, and legal systems that determine the validity and transferability of carbon units.

\subsection{Carbon Credits as Tokenized Real-World Assets}
Most modern financial and environmental assets are already represented electronically; tokenization therefore should not be understood merely as the conversion of an analogue record into digital form. Institutional analyses instead describe tokenization as the representation of an asset or claim on shared programmable infrastructure, potentially integrating ownership, transfer, servicing, settlement, and compliance logic \cite{BIS2024,Agur2025,Adrian2026}. Project Agorá provides a recent example from regulated financial infrastructure, demonstrating how programmable representations can coordinate conditional workflows and settlement without eliminating institutional controls \cite{BISAgora2026}.

Applied to carbon markets, this distinction shifts attention from the blockchain ledger itself to the relationship between the on-chain representation and the underlying carbon unit. Sorensen shows that tokenized carbon architectures differ substantially in how this relationship is constructed, including models based on retirement followed by token issuance, custodial or redeemable arrangements, and pooled representations of heterogeneous credits \cite{Sorensen2023}. The analysis also identifies a fundamental tension between fungibility and carbon-credit heterogeneity. Increasing fungibility may simplify trading and aggregation, but project type, methodology, vintage, permanence characteristics, and other attributes can remain economically and environmentally significant.

Empirical evidence also shows that tokenized carbon credits can retain links to the characteristics and value of the underlying asset. Swinkels examines carbon-credit tokens traded on AirCarbon Exchange and reports approximately 3.8 million tCO$_2$e of tokenized credits during 2021--2022, of which around 2.8 million tCO$_2$e were subsequently burned \cite{Swinkels2024}. Trading activity was concentrated in a small number of token categories. The most actively traded token also showed a close price relationship with comparable carbon credits in conventional markets. These findings indicate that tokenized carbon credits can preserve an economic connection to the underlying carbon asset.

These findings position carbon tokenization as a real-world-asset (RWA) representation problem. The environmental value originates upstream from mitigation activities, methodologies, measurement, reporting and verification (MRV), certification, and registry recognition. Tokenization operates downstream by changing how an already-recognized unit can be represented, transferred, programmed, and integrated with other applications. It cannot correct deficiencies in the environmental integrity of the underlying carbon credit.

\subsection{Public-Chain Carbon Markets and Empirical Evidence}

Public blockchain infrastructure extends tokenization by making token state and smart-contract execution available on a shared network that is not operated exclusively by a single carbon-market intermediary. Existing carbon-token projects demonstrate that such infrastructure can support trading, pooling, retirement, decentralized liquidity mechanisms, and integration with decentralized-finance applications. Ballesteros-Rodríguez et al.\ provide empirical evidence from KlimaDAO, showing that tokenized carbon credits can be incorporated into public-chain financial mechanisms while also documenting substantial changes in participation, staking activity, and token ownership over time \cite{Ballesteros2024}.

The growth of decentralized infrastructure does not remove the role of intermediaries or market concentration. Bassi et al.\ show that blockchain-based voluntary carbon markets remain dependent on influential actors such as tokenization providers, marketplaces, bridges, and registries \cite{Bassi2025}. Their findings indicate that blockchain changes the structure of intermediation but does not eliminate it. They also show that the use of a common blockchain does not necessarily lead to stronger economic interaction among market participants.

Evidence regarding liquidity is similarly cautious. Swinkels finds that operational carbon-token trading remained concentrated in relatively few instruments \cite{Swinkels2024}. Tlili reports no statistically significant increase in average daily trading volume following tokenization across the carbon-token platforms examined, although price responses vary with market and platform conditions \cite{Tlili2025}. More broadly, empirical evidence from tokenized RWA markets shows that token issuance and market capitalization can coexist with limited secondary-market turnover \cite{Mafrur2026}. Tokenization can therefore reduce technical barriers to transfer or market integration, but it should not be treated as a mechanism that mechanically produces liquidity.

Public-chain composability also introduces new risks. Tan argues that repeated transformation of carbon credits into pools, financial tokens, liquidity instruments, and other crypto-financial products can increase the distance between the traded instrument and the mitigation activity from which its environmental value originates \cite{Tan2024}. This critique is particularly important for carbon RWAs because the benefits of programmability and interoperability can be undermined if asset abstraction obscures project provenance or carbon-credit quality. The literature consequently supports a distinction between \emph{technical composability} and \emph{environmental integrity}: the former can be created by smart contracts, whereas the latter remains dependent on the underlying carbon system.

\subsection{Public and Permissioned Blockchain in Regulated Carbon Markets}

Public and permissioned blockchains differ across several dimensions, including network participation, validation, consensus, and governance \cite{NIST8202}. No single architecture is inherently suitable for all carbon-market applications. Permissioned and consortium blockchains can provide stronger access control, greater confidentiality, defined governance arrangements, and identifiable validators \cite{VilkovTian2023,Abiodun2025}. These characteristics are particularly relevant in regulated markets. For example, the K-ETS architecture proposed by Ajakwe et al.\ relies on authorized validators and institutional verification to support carbon-market operations \cite{Ajakwe2025}.

Public blockchain offers a different set of properties. Transaction and contract state can be independently observed, execution can be replicated across independently operated infrastructure, and standardized token interfaces can potentially be reused across multiple applications. These properties can reduce dependence on a single market operator for verification of the token-side state and can facilitate interoperability with other blockchain-based financial or carbon-management applications. They also introduce additional dependencies related to smart-contract security, public-network governance, privacy, transaction costs, and external infrastructure.

Crucially, base-layer openness does not require unrestricted access to the regulated asset itself. A public blockchain may permit broad observation and network validation while the carbon RWA contract restricts issuance, holding, or transfer to authorized participants. This distinction separates \emph{decentralization of transaction validation and execution} from \emph{authority over carbon issuance, participant eligibility, and environmental validity}. It therefore provides a conceptual basis for examining public blockchain within a regulated carbon-market setting without assuming anonymous or permissionless market participation.

\subsection{Registry Integrity and the Off-Chain--On-Chain Boundary}

The principal technical and institutional challenge in carbon tokenization arises at the boundary between the authoritative carbon unit and its blockchain representation. Blockchain can provide an immutable or tamper-evident history of token issuance and transfer, but it cannot independently establish whether the underlying mitigation activity is additional, correctly quantified, permanent, or otherwise environmentally credible \cite{VilkovTian2023,Tan2024}. The token therefore depends on information and institutional decisions originating outside the blockchain.

Carbon-market practice illustrates the importance of this relationship. Verra prohibited the creation of new crypto instruments based on already-retired Verified Carbon Units and subsequently considered approaches under which registry credits could be immobilized while corresponding tokenized representations circulated \cite{Verra2022}. Gold Standard has likewise developed conditions for third-party tokenization and explicitly identifies environmental-integrity, information-technology, regulatory, and reputational risks associated with such arrangements \cite{GoldStandard2024}. These developments reflect the broader principle identified by Sorensen: a tokenized carbon credit requires a credible mechanism linking the on-chain instrument to the carbon asset or methodology that substantiates it \cite{Sorensen2023}.

The same requirement follows from broader carbon-market integrity principles. The Integrity Council for the Voluntary Carbon Market emphasizes unique identification, registry tracking, independent verification, and safeguards against double issuance, double claiming, and double use \cite{ICVCM2024}. Public blockchain can make the token-side supply and transaction history more observable, but it cannot substitute for these upstream controls. A credible carbon RWA architecture must therefore ensure that the authoritative carbon unit and its tokenized representation do not become independently disposable claims on the same mitigation outcome.

This requirement also extends to the full asset lifecycle. Issuance, suspension, cancellation, transfer restrictions, and retirement can alter the authoritative status of the underlying carbon unit after token creation. Consequently, tokenization requires not only evidence that an eligible carbon unit existed at the time of minting, but also a mechanism through which subsequent changes in authoritative registry state can be reflected in the token state.

\subsection{Synthesis and Research Gap}

Several conclusions emerge from the reviewed literature. First, blockchain-enabled carbon trading, including peer-to-peer transfer and smart-contract execution, has already been extensively examined; the use of blockchain itself therefore cannot constitute the principal research contribution. Second, carbon credits have already been tokenized and traded in operational markets, demonstrating the feasibility of blockchain-based asset representation. Third, empirical evidence does not support the stronger proposition that tokenization automatically generates liquidity, broad participation, or decentralized market power. Fourth, public-chain composability introduces both opportunities and risks: it can expand programmability and interoperability while also increasing financial abstraction, governance dependencies, and the possibility that project-level attributes become obscured. Finally, the environmental validity of a carbon credit remains external to blockchain consensus and continues to depend on MRV, certification, registry governance, and lifecycle controls.

Across the literature reviewed here, however, two streams remain only partially connected. Research on regulated national emissions-trading systems has predominantly examined permissioned or hybrid blockchain architectures embedded within institutional governance, whereas empirical work on public-chain carbon tokens has largely emerged from voluntary carbon markets and crypto-financial ecosystems \cite{Ajakwe2025,Ballesteros2024,Bassi2025}. Existing work therefore provides limited guidance on how a carbon unit governed by a sovereign national registry can be represented as a portable RWA on public blockchain infrastructure while preserving authoritative registry control, project provenance, controlled issuance, participant eligibility, synchronization of asset states, and authoritative retirement.

This gap is particularly relevant to Indonesia because the research problem is not the
introduction of digital carbon trading into an otherwise analogue market. Indonesia already
operates a regulated carbon-market framework and blockchain-enabled trading infrastructure.
The unresolved design question is therefore more specific: how can a public-chain RWA layer
provide additional programmability, independent verifiability, and interoperable asset
representation without becoming a competing source of carbon validity or bypassing the existing
regulatory and registry architecture?

The architecture developed in this study is derived from three sources of evidence. Indonesian
regulations and institutional documentation establish the roles of SRUK, Carbon Exchange
trading, direct trading, and regulatory authorities. Current IDXCarbon evidence establishes the
existing digital market baseline. The carbon-tokenization and tokenized-finance literature
identifies requirements associated with provenance, controlled issuance, participant eligibility,
registry synchronization, and retirement. Together, these sources lead to six principal design
requirements: registry primacy, controlled token issuance, preservation of provenance,
lifecycle synchronization, participant eligibility, and authoritative retirement. These requirements inform the architecture developed in Section~\ref{sec:architecture}.

\section{Why Tokenize Carbon Credits and Why Public Blockchain?}
\label{sec:why_tokenize}

Indonesian carbon credits do not require tokenization in order to become digital. SPE-GRK is already electronically recorded and can be traded through established market mechanisms \cite{IDXSPE,IDXFAQ}. Conventional databases can also provide transaction processing, access controls, audit records, and application programming interfaces. The rationale for tokenization must therefore be based on capabilities that arise from representing the carbon unit on programmable infrastructure.

\subsection{From Digital Carbon Credits to Programmable RWAs}

Tokenization introduces an additional representation of an existing carbon unit on a shared programmable ledger. The IMF distinguishes tokenization from conventional digitization by emphasizing the representation of assets and related transactions within programmable infrastructure \cite{Agur2025}. Applied to Indonesian carbon credits, the relationship can be expressed as

\[
\underbrace{\text{SPE-GRK in authoritative registry}}_{\text{digital carbon unit}}
\quad\longrightarrow\quad
\underbrace{\text{public-chain RWA token}}_{\text{programmable representation}}.
\]

The token does not replace the SPE-GRK or determine its environmental validity. It represents a carbon unit whose status remains governed by the authoritative registry. The legal and economic rights attached to this representation must therefore be clearly defined.

One potential benefit of tokenization is programmable lifecycle management. Carbon credits move through several states, including issuance, holding, transfer, pooling, and retirement. Smart contracts can represent these transitions and apply predefined conditions to them. For example, a corporate emissions-management application could acquire an eligible token and initiate a retirement process. Settlement could be conditional on payment and participant eligibility. Portfolio applications could also restrict purchases to credits that satisfy specified project, methodology, or vintage criteria.

Tokenization can also provide a common interface for interacting with carbon assets across different applications. Conventional carbon systems typically record units within accounts maintained by registries or market operators. A token introduces an additional wallet-based representation. NIST notes that blockchain token architectures can support digital asset control through public-key infrastructure, subject to the custody arrangement adopted \cite{NIST8301}. Regulated implementations can use institutional wallets, qualified custodians, multisignature arrangements, and account-recovery mechanisms.

Standardized token interfaces further support application interoperability. ERC-20 provides a common interface for fungible tokens, whereas ERC-1155 allows multiple token classes to be represented within a single contract \cite{ERC20,ERC1155}. The latter is relevant to carbon markets because credits remain heterogeneous across projects, vintages, methodologies, geography, and authorization status. An ERC-1155-style structure can preserve these distinctions at the batch level. Fungible pooling may subsequently be introduced where the underlying credits satisfy a clearly defined eligibility policy.

Standardization can simplify integration with wallets, carbon-accounting applications, portfolio systems, settlement infrastructure, and supervisory tools. Related initiatives such as the Climate Action Data Trust demonstrate the broader value of standardized data interfaces for connecting independently governed carbon registries \cite{WorldBankCAD}. Tokenization extends this principle from data exchange to programmable asset representation.

Programmability can also support settlement coordination. Where a legally permitted settlement asset is available on compatible infrastructure, delivery-versus-payment conditions can be incorporated into smart-contract workflows. Project Agorá illustrates how programmable infrastructure can combine transaction conditions, compliance checks, and tokenized settlement assets in a regulated financial setting \cite{BISAgora2026}. An Indonesian carbon RWA architecture should nevertheless remain neutral with respect to the settlement instrument. Payment may continue to use conventional banking infrastructure or another mechanism permitted under applicable regulation.

\subsection{The Case for Public Blockchain}

Tokenization does not require a public blockchain. Private and consortium networks can support smart contracts, controlled participation, rapid settlement, and transaction transparency within the network. They may also provide stronger confidentiality and more direct governance control.

The additional value of a public blockchain lies in providing a shared execution and verification environment beyond a single market operator. Token supply, transfers, contract activity, and lifecycle events can be independently observed from the blockchain. Standardized asset interfaces can also be accessed by multiple authorized applications using the same underlying token representation. This distinction is particularly relevant where several institutions need to interact with the same carbon asset. A corporate carbon-management platform, regulated trading venue, custodian, settlement provider, auditor, and supervisory application could use a common token interface while continuing to perform different institutional functions. Some interoperability requirements can be addressed through conventional APIs. A public-chain RWA becomes more relevant when multiple systems need to transact against a common asset state and execute shared transaction logic.

\begin{table}[!htbp]
\centering
\caption{Comparison of existing exchange-centred infrastructure and a public-chain RWA layer}
\label{tab:comparison}

\footnotesize
\setlength{\tabcolsep}{3pt}
\renewcommand{\arraystretch}{0.92}

\begin{tabularx}{\textwidth}{
    >{\raggedright\arraybackslash}p{2.5cm}
    >{\raggedright\arraybackslash}X
    >{\raggedright\arraybackslash}X}
\toprule
\textbf{Dimension} &
\textbf{Existing exchange-centred model} &
\textbf{Public-chain RWA model} \\
\midrule

Carbon trading &
Available through regulated mechanisms. &
Adds a public-chain representation; it does not create the ability to trade carbon. \\

Asset representation &
Carbon units remain within registry and market infrastructure. &
An additional standardized token represents the carbon unit on public blockchain. \\

Programmability &
Rules can be implemented within operator-controlled applications. &
Lifecycle and transfer rules can be incorporated into the token interface. \\

Interoperability &
Integration generally relies on institution-specific interfaces. &
Standardized token interfaces can support multiple authorized applications. \\

Wallet interaction &
Access depends on institutional account and custody arrangements. &
Authorized wallets can interact with the token through compatible applications. \\

Observability &
Information depends on registry, exchange, and institutional reporting. &
Token supply, transfers, and contract events can be independently inspected. \\

Participant eligibility &
Determined by applicable market and institutional rules. &
Eligibility controls can be incorporated into token-transfer rules. \\

Environmental integrity &
Depends on MRV, verification, methodology, and registry governance. &
Remains dependent on the same upstream carbon-integrity mechanisms. \\

Liquidity &
Depends on participants, asset characteristics, and market structure. &
Tokenization does not itself create secondary-market liquidity. \\

Additional risks &
Operational, market, cyber, and institutional risks. &
Adds smart-contract, wallet, public-network, privacy, and interface risks. \\

\bottomrule
\end{tabularx}
\end{table}

Table~\ref{tab:comparison} summarizes the main differences between the existing
exchange-centred model and the proposed public-chain RWA layer. The public blockchain therefore provides an additional infrastructure layer for asset representation, verification, and programmable interaction. Its value depends on whether these capabilities justify the additional technical, governance, privacy, and regulatory requirements associated with operating on public infrastructure.

\subsection{Limits and Conditions of Public-Chain Tokenization}

The benefits of tokenization should be separated from claims about carbon integrity and market performance. Public blockchain can make the state of a token independently observable, but this does not establish the environmental quality of the underlying carbon credit. Blockchain records can show that a token was issued, transferred, frozen, or burned according to predefined rules. They cannot establish additionality, permanence, baseline accuracy, leakage, or the reliability of the underlying MRV process \cite{ICVCM2024,Tan2024}. These remain functions of carbon methodologies, verification processes, and registry governance.

The same distinction applies to liquidity. Tokenization is frequently associated with expectations of improved market access and secondary trading, but existing evidence does not support an automatic relationship between tokenization and liquidity. Swinkels finds that trading in operational carbon-token markets remained concentrated in a limited number of instruments \cite{Swinkels2024}. Tlili reports limited effects on short-term transaction volumes across the carbon-token markets examined \cite{Tlili2025}. Similar heterogeneity is observed across other tokenized RWA markets \cite{Mafrur2026}.

Tokenization can therefore create new market capabilities, including standardized asset representation, programmable transactions, wallet interoperability, and broader application integration. The effect of these capabilities on participation and liquidity depends on market demand, asset quality, pricing, transaction costs, market-making activity, and regulatory conditions. These outcomes require empirical evaluation and are not assumed in the proposed architecture.

\section{Architecture for Public-Blockchain RWA Carbon Trading in Indonesia}
\label{sec:architecture}

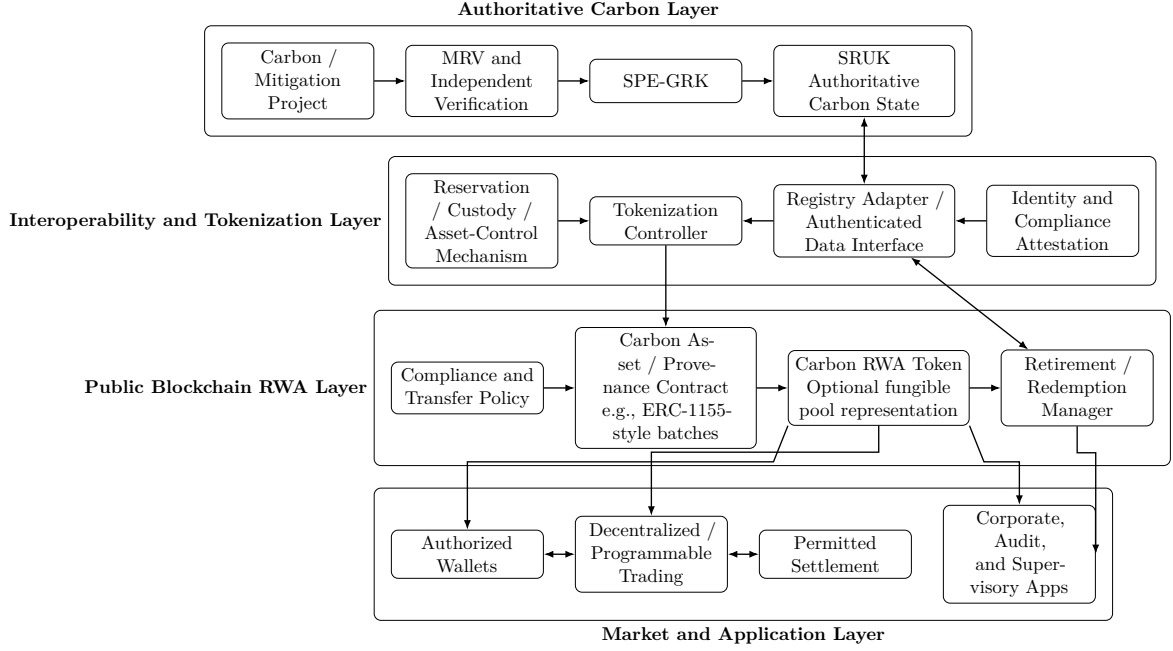
\begin{figure}[htbp]
\centering
\resizebox{0.98\textwidth}{!}{
\begin{tikzpicture}[
    node distance=0.65cm and 0.65cm,
    box/.style={
        draw,
        rounded corners,
        align=center,
        minimum height=0.9cm,
        text width=2.9cm
    },
    widebox/.style={
        draw,
        rounded corners,
        align=center,
        minimum height=0.9cm,
        text width=3.5cm
    },
    arrow/.style={-{Latex[length=2mm]}, thick},
    biarrow/.style={{Latex[length=2mm]}-{Latex[length=2mm]}, thick},
    group/.style={draw, rounded corners, inner sep=0.35cm}
]


\node[box] (project) {Carbon / Mitigation\\Project};
\node[box, right=of project] (mrv) {MRV and Independent\\Verification};
\node[box, right=of mrv] (spe) {SPE-GRK};
\node[widebox, right=of spe] (sruk) {SRUK\\Authoritative Carbon State};


\node[widebox, below=1.4cm of sruk]
    (adapter)
    {Registry Adapter /\\Authenticated Data Interface};

\node[box, left=of adapter]
    (tokenizer)
    {Tokenization\\Controller};

\node[box, left=of tokenizer]
    (custody)
    {Reservation / Custody /\\Asset-Control Mechanism};

\node[box, right=of adapter]
    (identity)
    {Identity and\\Compliance Attestation};


\node[widebox, below=1.7cm of tokenizer]
    (carbonbatch)
    {Carbon Asset / Provenance Contract\\
     e.g., ERC-1155-style batches};

\node[widebox, right=of carbonbatch]
    (rwa)
    {Carbon RWA Token\\
     Optional fungible pool representation};

\node[box, left=of carbonbatch]
    (policy)
    {Compliance and\\Transfer Policy};

\node[box, right=of rwa]
    (retire)
    {Retirement /\\Redemption Manager};


\node[box, below=2.4cm of policy]
    (wallet)
    {Authorized\\Wallets};

\node[box, right=of wallet]
    (trade)
    {Decentralized /\\Programmable Trading};

\node[box, right=of trade]
    (settle)
    {Permitted\\Settlement};

\node[box, right=of settle]
    (apps)
    {Corporate, Audit,\\and Supervisory Apps};


\draw[arrow] (project) -- (mrv);
\draw[arrow] (mrv) -- (spe);
\draw[arrow] (spe) -- (sruk);

\draw[biarrow] (sruk) -- (adapter);
\draw[arrow] (adapter) -- (tokenizer);
\draw[arrow] (custody) -- (tokenizer);
\draw[arrow] (identity) -- (adapter);

\draw[arrow] (tokenizer) -- (carbonbatch);
\draw[arrow] (carbonbatch) -- (rwa);
\draw[arrow] (policy) -- (carbonbatch);
\draw[biarrow] (adapter) -- (retire);
\draw[arrow] (rwa) -- (retire);

\draw[biarrow] (wallet) -- (trade);
\draw[biarrow] (trade) -- (settle);

\draw[arrow] (rwa.south) -- ++(0,-0.55) -| (trade.north);
\draw[arrow] (rwa.south west) -- ++(-0.3,-0.75) -| (wallet.north);
\draw[arrow] (rwa.south east) -- ++(0.3,-0.75) -| (apps.north);

\draw[arrow] (retire.south) -- ++(0,-0.55) -| (apps.east);


\node[
    group,
    fit=(project)(mrv)(spe)(sruk),
    label={[font=\bfseries]above:Authoritative Carbon Layer}
] {};

\node[
    group,
    fit=(custody)(tokenizer)(adapter)(identity),
    label={[font=\bfseries]left:Interoperability and Tokenization Layer}
] {};

\node[
    group,
    fit=(policy)(carbonbatch)(rwa)(retire),
    label={[font=\bfseries]left:Public Blockchain RWA Layer}
] {};

\node[
    group,
    fit=(wallet)(trade)(settle)(apps),
    label={[font=\bfseries]below:Market and Application Layer}
] {};

\end{tikzpicture}
}

\caption{Proposed architecture for public-blockchain RWA carbon trading in Indonesia. SRUK maintains the authoritative carbon-unit state, while the public-blockchain layer supports token representation and programmable transactions.}
\label{fig:proposedarchitecture}
\end{figure}

Figure~\ref{fig:proposedarchitecture} presents the proposed architecture. It comprises four logical layers: the authoritative carbon layer, the interoperability and tokenization layer, the public-blockchain RWA layer, and the market and application layer. The layers are connected through controlled interfaces so that changes in the token state remain consistent with the corresponding carbon-unit state.

\subsection{Architecture Overview}

The authoritative carbon layer contains the processes that establish and maintain the validity of the carbon unit. These include mitigation-project registration, measurement, reporting and verification (MRV), independent verification, issuance, registry recording, status changes, and retirement recognition. Tokenization begins only after an eligible carbon unit has been recognized within the applicable Indonesian framework.

Each tokenized carbon asset must retain a resolvable relationship with its underlying carbon unit. Relevant attributes include the registry identifier, project, quantity, vintage, methodology, current status, and retirement state. These attributes do not all need to be stored directly on a public blockchain. The on-chain record may contain identifiers or cryptographic commitments that reference information maintained within SRUK or another authorized system.

This separation preserves the institutional role of the carbon registry. Blockchain consensus can verify token transactions and execute contract rules, but the validity of the represented carbon unit continues to depend on the authoritative carbon-market system.

\subsection{Registry Integration and Tokenization Control}

The interoperability and tokenization layer connects SRUK with the public-chain representation. A registry adapter provides authenticated information on the identity, quantity, and status of carbon units selected for tokenization. It must also communicate subsequent events that affect the tokenized asset, including suspension, cancellation, or retirement. An authenticated machine-to-machine interface would provide the most direct form of integration. If such an interface is unavailable, an authorized gateway could relay signed registry information. The architecture does not assume that a production SRUK API or equivalent interface is currently available. The availability and governance of this interface are therefore implementation requirements.

Token issuance also requires a mechanism that prevents the same carbon quantity from remaining independently transferable through both the registry system and the blockchain representation. A registry-native tokenization or immobilization state would provide a direct solution. Other implementations could use segregated custody or a legally enforceable reservation mechanism.

For a carbon batch \(b\), the following condition should hold:

\begin{equation}
T_b \leq Q^{\mathrm{reserved}}_b,
\label{eq:supply}
\end{equation}

where \(T_b\) denotes the quantity represented by live tokens and
\(Q^{\mathrm{reserved}}_b\) denotes the quantity reserved for tokenization.

This condition ensures that token issuance remains bounded by the carbon quantity committed to the tokenized representation. Confirmation that a carbon credit exists is insufficient if the same unit can still be transferred independently outside the tokenization arrangement.

\subsection{Public-Chain Asset Representation and Compliance}

The public-blockchain layer represents the carbon asset, its provenance, and the rules governing its transfer. Carbon heterogeneity should be preserved at this stage. Project, vintage, methodology, authorization status, and other attributes can affect both the economic and environmental characteristics of a carbon credit. An ERC-1155-style structure provides one possible implementation because multiple carbon batches can be represented within the same contract using separate token identifiers \cite{ERC1155}. Each identifier can correspond to a defined batch or class and retain a reference to the underlying registry information. Sensitive project or participant information can remain off-chain, provided that sufficient information is available to verify provenance and reconcile the token with the authoritative record.

Fungible representations may also be useful for standardized carbon products. An ERC-20-style token could represent a pool of eligible credits where a defined policy establishes sufficient comparability among the underlying assets \cite{ERC20}. Pooling must be designed carefully because aggregation can obscure differences between projects and credit characteristics \cite{Sorensen2023}. The composition of a pool should therefore remain traceable to the underlying carbon units.

Participant controls can be incorporated into the same infrastructure. Public blockchain validation does not require unrestricted access to the carbon asset. Wallets interacting with the RWA contract can be linked to verified participant credentials, and smart contracts can enforce eligibility requirements before permitting transfers. Personal identity information can remain off-chain, with the blockchain storing only the status needed to determine whether an address is authorized. This model is consistent with developments in regulated tokenized finance, where programmable infrastructure is increasingly combined with identity and compliance controls \cite{BIS2024,Adrian2026}. Public visibility applies to the token and its transactions; access to the regulated carbon asset remains subject to the applicable market rules.

\subsection{Trading, Settlement, Retirement, and Supervision}

The tokenized asset could operate under several trading arrangements, depending on the
applicable regulatory classification. One possible implementation path is to use the public
blockchain for asset representation and transaction execution within authorized direct-trading
arrangements, with buyer and seller matching remaining bilateral or occurring through an
approved interface. The tokenized representation could also be integrated with a licensed
Carbon Exchange, allowing IDXCarbon or another authorized operator to continue providing
market functions while the public blockchain supports asset representation and transaction
execution. More decentralized mechanisms, including multilateral smart-contract protocols or
automated liquidity arrangements, are technically possible but would require separate regulatory
assessment. The architecture therefore does not assume that permissionless decentralized
exchanges or automated market makers are permitted for Indonesian carbon units.

Settlement should remain independent of the choice of trading mechanism. Where payment
continues through conventional banking infrastructure, smart contracts could coordinate asset
transfer with external payment confirmation or an escrow arrangement. More direct
delivery-versus-payment execution may become possible if an approved digital settlement
instrument becomes available. The architecture does not prescribe cryptocurrency or stablecoin
settlement because the permitted payment mechanism depends on the applicable Indonesian
financial and payment framework.

Retirement requires closer integration with SRUK because it changes the authoritative status
of the underlying carbon unit. Burning a token on the public blockchain is therefore insufficient
to establish retirement. The proposed process is

\[
\text{Retirement Request}
\rightarrow
\text{Token Freeze}
\rightarrow
\text{SRUK Confirmation}
\rightarrow
\text{Token Burn}.
\]

Following a retirement request, the token is first made non-transferable. The corresponding
retirement is then processed through the authoritative registry, and the token is burned only
after confirmation from SRUK. If the retirement is rejected or cannot be confirmed, the token
must remain subject to a defined reconciliation process. This sequence prevents the blockchain
from recording final retirement before the underlying carbon unit has been retired within the
authoritative system.

The public blockchain also provides an additional source of information for supervision and
audit. Token issuance, supply, transfers, retirement requests, burn events, contract upgrades,
and administrative actions can be observed directly from blockchain activity. These records
complement the authoritative information maintained by SRUK rather than replace it. SRUK
determines the legal and regulatory status of the carbon unit, and the public blockchain records
the state and transaction history of its tokenized representation. Maintaining consistency between
these two records is therefore a central operational requirement of the proposed architecture.

\section{Discussion}
\label{sec:discussion}

The findings show that the contribution of public-blockchain tokenization in Indonesia does not arise from the ability to trade carbon or from the use of blockchain technology itself. Both already exist within the current market infrastructure. The value of the proposed model lies in extending the representation of the carbon unit beyond an exchange-specific environment and making that representation available through a common programmable interface.

\subsection{Public-Chain Contribution and the Proposed Architecture}

RQ1 examines the additional benefits and challenges that public-blockchain tokenization may introduce into Indonesia's existing carbon market. The analysis identifies three main capabilities. First, tokenization provides a standardized representation of the carbon unit that can carry lifecycle and transfer rules through smart contracts. Second, public blockchain makes the token-side state independently observable. Supply, transfers, contract activity, and retirement events can be inspected without relying exclusively on an interface operated by the token issuer or market platform. Third, standardized token interfaces can allow the same asset representation to interact with multiple authorized applications.

These capabilities do not imply that a public blockchain is superior for every carbon-market function. Private infrastructure can provide high transaction performance, controlled access, confidentiality, and programmable logic. Some interoperability requirements can also be addressed through conventional APIs. The public-chain model becomes more relevant when several institutions need to interact with the same asset state. Examples include trading platforms, custodians, corporate carbon-management systems, settlement providers, auditors, and supervisory applications. The empirical literature also places limits on the expected market effects. Tokenization does not automatically create liquidity. Swinkels reports concentrated trading activity in operational carbon-token markets \cite{Swinkels2024}, while Tlili finds limited short-term effects on transaction volume \cite{Tlili2025}. Similar variation is observed across other tokenized RWA markets \cite{Mafrur2026}. The economic value of tokenization therefore depends on actual demand, asset quality, pricing, transaction costs, and market structure.

RQ2 concerns the architecture required to support this form of tokenization. The proposed model preserves SRUK as the authoritative source of carbon-unit state. The public blockchain records the corresponding token state and executes approved transaction rules. An interoperability layer connects the two systems and communicates information on asset identity, eligible quantity, status, and retirement. This arrangement creates a clear division of responsibility. SRUK determines whether a carbon unit exists and remains valid. The public blockchain records how its tokenized representation is issued, transferred, restricted, and retired. The tokenization mechanism must maintain consistency between these two states. A carbon unit committed to tokenization cannot remain independently transferable through another channel. The same principle applies to retirement. Burning a blockchain token alone does not establish that the environmental claim has been retired. The authoritative registry must first confirm the change in carbon-unit status. The proposed sequence therefore freezes the token after a retirement request, obtains confirmation from SRUK, and then completes the on-chain retirement.

Public blockchain also does not require unrestricted participation. The network may remain open for validation and observation while access to the carbon asset is limited to verified participants. Identity information can remain off-chain. Smart contracts only need sufficient information to determine whether an address is eligible to hold or transfer the asset. This model is consistent with the broader development of regulated tokenized finance \cite{BIS2024,Adrian2026}. The resulting architecture is therefore hybrid in institutional terms. Carbon validity remains within the sovereign carbon-market framework. Public blockchain provides a common asset and execution layer. Market participation remains subject to Indonesian regulation.

\subsection{Implementation in Indonesia}

RQ3 concerns the conditions required to implement the proposed architecture in Indonesia.
The main requirements are reliable integration with SRUK, legal and regulatory clarity,
accountable governance, and identifiable market use cases. At the present stage, these issues
are more fundamental than transaction throughput or blockchain performance.

A production implementation would first require a reliable interface with SRUK. The system
must confirm the identity, eligible quantity, and current status of a carbon unit before token
issuance and receive subsequent changes affecting transferability, suspension, cancellation, or
retirement. Authentication, reconciliation, outage handling, and protection against replay or
fraudulent messages would therefore be required. Legal and regulatory treatment is equally important. POJK No.\ 10 of 2026 explicitly
classifies a Unit Karbon as an Efek \cite{OJK10}. Tokenization does not by itself alter the
regulatory status of the underlying carbon unit. The unresolved issue concerns the legal
character of its on-chain representation and the relationship between token transfer and the
rights attached to the underlying SPE-GRK. Possible structures may involve legal title,
beneficial interest, or a contractual claim, each with different implications for custody,
transfer, settlement, and investor protection. Trading functions would also require regulatory classification. Perpres 110/2025 recognizes
Carbon Exchange and direct-trading channels \cite{Perpres110}, and Carbon Exchange activity
is regulated under POJK No.\ 14 of 2023 and POJK No.\ 10 of 2026
\cite{OJK14,OJK10}. A public-chain platform performing multilateral market functions would
need to be assessed according to the activity it performs. POJK No.\ 23 of 2025 provides a
separate regulatory framework for Digital Financial Assets using distributed-ledger technology
\cite{OJK23}. Whether, and to what extent, this framework would apply to an on-chain
representation of SPE-GRK would require regulatory determination before production deployment.

Governance and market justification are equally important. Token issuance, asset eligibility,
contract upgrades, emergency suspension, participant authorization, and registry connectivity
require clearly assigned institutional responsibilities. These functions should not depend on a
single administrator or private key. Responsibilities could be distributed across the carbon
registry authority, token issuer, compliance provider, market operator, and smart-contract
administration process. The additional infrastructure must also address identifiable market
needs because IDXCarbon already provides trading, project listing, participant registration,
and retirement functions. Relevant use cases include automated corporate procurement,
programmable retirement, integration with carbon-accounting systems, coordinated settlement,
and a common carbon-asset interface across authorized applications. OJK's regulatory-sandbox
experience provides a possible environment for testing these arrangements. Tokenization models
involving gold, securities, and property had completed sandbox processes by March 2026
\cite{OJKSandbox2026}, providing a relevant precedent for controlled experimentation with
tokenized assets.

A practical implementation pathway should therefore begin with a limited pilot involving a
small quantity of SPE-GRK and a restricted group of verified participants. The pilot should test
the complete lifecycle from registry confirmation and token issuance to transfer, settlement,
retirement, and reconciliation with SRUK. It should also evaluate identity controls, custody,
the legal effect of token transfers, and smart-contract governance. Integration with authorized
direct trading or IDXCarbon could follow once the domestic token lifecycle has been demonstrated
to operate reliably. Broader interoperability should be introduced more cautiously because
international connectivity, cross-chain bridges, and multiple token representations increase
legal, technical, and reconciliation risks. Early interoperability should therefore focus on
common metadata, APIs, and standardized asset interfaces.

\subsection{Risks and Implications}

The proposed architecture introduces risks at both the carbon-asset and infrastructure levels.
The first concerns environmental integrity and provenance. A public blockchain can provide an
auditable history of token issuance, transfer, and retirement, but it cannot establish the quality
of the underlying mitigation outcome. Additionality, permanence, baseline accuracy, leakage,
and MRV quality remain functions of the carbon-market system
\cite{ICVCM2024,Tan2024}. Public-chain financialization can also increase the distance between
the traded instrument and the underlying mitigation project. Carbon tokens may be pooled,
wrapped, collateralized, or incorporated into other digital-financial products, potentially
obscuring differences in project type, vintage, methodology, and environmental characteristics
\cite{Tan2024,Sorensen2023}. The architecture should therefore preserve project- and
batch-level provenance even when standardized or pooled representations are introduced.

A related risk concerns fragmentation and duplicate representation. The same underlying carbon
unit could potentially be represented through multiple tokens, pools, platforms, or blockchain
networks. This would increase reconciliation requirements and create uncertainty over which
representation is authoritative or transferable. Tokenization should therefore follow a controlled
issuance process linked to the corresponding registry state. Additional representations should
only be created when the underlying token or carbon quantity is appropriately locked, reserved,
or otherwise prevented from circulating independently. Cross-chain replication should remain
limited until reliable mechanisms exist for maintaining consistency across representations.

Public-chain deployment also expands the technical risk surface. Smart-contract vulnerabilities,
compromised administrative keys, wallet loss, network congestion, and failures in external
services can affect the operation of the tokenized asset. The registry interface is particularly
important because the public blockchain depends on authenticated information from SRUK.
The objective is not to reproduce carbon-state determination on-chain, but to communicate
authoritative registry changes securely and consistently to the relevant contracts. Production
deployment would therefore require secure authentication, reconciliation procedures, monitored
contract upgrades, separation of administrative responsibilities, emergency controls, and
mechanisms for handling unavailable or conflicting registry information.

These risks reinforce a broader implication of the proposed architecture. Token creation is not
the most difficult part of carbon-market tokenization. The more significant challenges concern
the rights represented by the token, authority over issuance, consistency with SRUK, participant
eligibility, settlement, retirement, governance, and resolution of conflicts between on-chain and
authoritative records. Public-blockchain carbon RWAs should therefore be evaluated as an
extension of Indonesia's existing carbon-market infrastructure. Their value depends on whether
programmable and interoperable asset functions provide sufficient practical benefit to justify the
additional technical, regulatory, governance, and reconciliation requirements.

\subsection{Limitations and Future Research}

The proposed architecture is conceptual and several implementation questions remain unresolved. 
The study does not have access to a production SRUK technical interface, and public documentation 
does not establish whether functions such as APIs, asset reservation, or registry-native 
immobilization are currently available. The legal status of a tokenized SPE-GRK also requires 
clarification, including the rights represented by the token, custody, transfer finality, dispute 
resolution, and the regulatory treatment of public-chain trading functions 
\cite{Perpres110,OJK14,OJK10}. Technical choices concerning the blockchain network, privacy, 
identity, registry integration, metadata, and governance would likewise need to be evaluated 
before production deployment.

The study also does not estimate the economic or environmental effects of public-chain 
tokenization. IDXCarbon data provide market context but cannot establish whether the proposed 
architecture would increase trading activity, improve price discovery, reduce transaction costs, 
or attract new participants. Tokenization also does not alter the underlying quality of the carbon 
credit, which remains dependent on additionality, permanence, MRV, methodology, and verification. 
Future research should therefore evaluate the architecture through prototypes, stakeholder 
interviews, and supervised pilots covering issuance, transfer, settlement, suspension, and 
retirement. Such evaluation could compare existing and tokenized workflows using measures such 
as settlement time, reconciliation effort, operational cost, auditability, and market participation.

\section{Conclusion}

Indonesia already operates a regulated and digitally enabled carbon market, with SRUK providing the authoritative registry function and IDXCarbon supporting blockchain-based trading. This study therefore focuses on the additional role that public-blockchain tokenization could play within the existing market structure. The proposed architecture represents SPE-GRK as a tokenized RWA while preserving SRUK as the authoritative source of carbon-unit status. Public blockchain is used to support programmable asset representation, transaction execution, interoperability, and independent observation of token activity. Access to the tokenized asset remains subject to regulatory, identity, and registry controls. Overall, public-blockchain carbon tokenization is best understood as an extension of Indonesia's existing carbon-market infrastructure. Its potential lies in adding programmable and interoperable asset functions while maintaining the institutional controls required for credible carbon trading.

\appendix
\section{Abbreviations and Key Terms}
\label{app:terms}

\setcounter{table}{0}
\renewcommand{\thetable}{A\arabic{table}}

\begin{table}[H]
\centering
\caption{Abbreviations and key terms used in this study}
\label{tab:terms}

\footnotesize
\setlength{\tabcolsep}{4pt}
\renewcommand{\arraystretch}{1.05}

\begin{tabularx}{\textwidth}{
>{\raggedright\arraybackslash}p{3.0cm}
>{\raggedright\arraybackslash}X}
\toprule
\textbf{Term} & \textbf{Meaning} \\
\midrule

API &
\textit{Application Programming Interface}. A technical interface that allows
different software systems to exchange data or instructions. In the proposed
architecture, an API could be used to communicate registry information to other
authorized systems. \\

Carbon credit / carbon unit &
A recognized unit representing a quantified amount of greenhouse-gas emission
reduction or removal. Carbon units remain subject to the rules of the relevant
carbon-market and registry framework. \\

Carbon Exchange &
A regulated market infrastructure through which eligible carbon units can be
traded. IDXCarbon is the Carbon Exchange considered in this study. \\

ERC-20 &
An Ethereum token standard commonly used for fungible tokens, where units of the
same token are generally interchangeable. In this study, it is considered as a
possible representation for standardized pools of eligible carbon credits. \\

ERC-1155 &
An Ethereum multi-token standard that allows multiple token classes to be managed
within a single smart contract. It is relevant to carbon credits because different
projects, vintages, or credit categories can be represented using separate token
identifiers. \\

IDXCarbon &
Indonesia's Carbon Exchange operated by PT Bursa Efek Indonesia. It provides
regulated mechanisms for trading eligible Indonesian carbon units. \\

Mitigation &
An activity that reduces or avoids greenhouse-gas emissions, or increases greenhouse-gas
removals. Examples include renewable-energy projects, methane capture, energy-efficiency
projects, and forest restoration. \\

Mitigation project &
A project or activity that generates measurable greenhouse-gas reductions or removals.
The existence of a project does not itself create carbon credits; the mitigation
outcome must satisfy the applicable measurement, verification, and regulatory
requirements. \\

MRV &
\textit{Measurement, Reporting, and Verification}. The processes used to quantify,
document, and verify greenhouse-gas emission reductions or removals associated
with a mitigation activity. \\

NEK &
\textit{Nilai Ekonomi Karbon} (Carbon Economic Value). Indonesia's framework for
the economic valuation and implementation of carbon-related instruments. \\

OJK &
\textit{Otoritas Jasa Keuangan} (Financial Services Authority of Indonesia).
OJK regulates and supervises Carbon Exchange activity in Indonesia. \\

Public blockchain &
A blockchain network whose transaction and smart-contract state can generally be
independently observed and validated through publicly accessible network
infrastructure. Public blockchain does not necessarily imply unrestricted access
to the regulated carbon asset. \\

Retirement &
The process through which a carbon unit is permanently removed from further use
or trading after being applied to an eligible carbon or emissions claim. A retired
unit cannot subsequently be sold or used again. \\

RWA &
\textit{Real-World Asset}. An asset or claim whose economic or legal basis exists
outside the blockchain but is represented on programmable blockchain
infrastructure. In this study, the underlying asset is an Indonesian carbon unit. \\

SPE-GRK &
\textit{Sertifikat Pengurangan Emisi Gas Rumah Kaca}. An Indonesian certificate
representing recognized greenhouse-gas emission reductions under the applicable
national carbon-market framework. \\

SRUK &
\textit{Sistem Registri Unit Karbon}. Indonesia's national carbon-unit registry.
In the proposed architecture, SRUK remains the authoritative source for the
existence and status of the underlying carbon unit. \\

Tokenization &
The representation of an existing asset or claim as a token on programmable
blockchain infrastructure. Tokenization does not create the underlying carbon
credit or establish its environmental validity. \\

tCO$_2$e &
Tonnes of carbon-dioxide equivalent. A common unit used to express the climate
impact of different greenhouse gases on a comparable CO$_2$-equivalent basis. \\

Vintage &
The period, commonly a year, in which the underlying emission reduction or removal
occurred. Carbon credits from different vintages may have different market or
environmental characteristics. \\

\bottomrule
\end{tabularx}
\end{table}



\end{document}